\documentstyle[12pt]{article}

\newcommand{\eq}{\begin{equation}}
\newcommand{\en}{\end{equation}}
\newcommand{\eqa}{\begin{eqnarray}}
\newcommand{\ena}{\end{eqnarray}}

\newcommand{\lhi}{\hat\lambda_{i}}

\begin{document}

\hskip 11.5cm \vbox{\hbox{DFTT 38/92}\hbox{July 1992}}
\vskip 0.4cm
\centerline{\bf   EXACT SOLUTION OF D=1 KAZAKOV-MIGDAL}
\centerline{\bf INDUCED GAUGE THEORY}
\vskip 1.3cm
\centerline{ M. Caselle, A. D'Adda  and  S. Panzeri}
\vskip .6cm
\centerline{\sl  Dipartimento di Fisica
Teorica dell'Universit\`a di Torino}
\centerline{\sl Istituto Nazionale di Fisica Nucleare,Sezione di Torino}
\centerline{\sl via P.Giuria 1, I-10125 Turin,Italy}
\vskip 2.5cm

\begin{abstract}
We give the exact solution of the Kazakov-Migdal induced gauge model
in the case of a D=1 compactified lattice with a generic number $S$ of
sites and for any value of N. Due to the peculiar features of the model,
the partition function that we obtain also describes the vortex-free
sector of the D=1 compactified bosonic string, and it coincides in the
continuum limit with the one obtained by Boulatov and Kazakov in this
context.
\end{abstract}
\vskip 2.5cm
\hrule
\vskip1.2cm
\noindent

\hbox{\vbox{\hbox{$^{\diamond}${\it email address:}}\hbox{}}
 \vbox{\hbox{ Decnet=(31890::CASELLE,DADDA,PANZERI)}
\hbox{ Bitnet=CASELLE(DADDA)(PANZERI)@TORINO.INFN.IT}}}
\vfill
\eject

\newpage

{\bf 1. Introduction}
\vskip .3cm

Recently, a new interesting approach to  gauge theories has
been proposed~\cite{km} and solved in the large N limit~~\cite{km,migdal}
 by Kazakov and Migdal. The hope is to be able to describe within this
approach, the
asymptotically free fixed point of QCD in 4 dimensions. The
Kazakov-Migdal model seem to be a promising tool in this direction, but
several question must be understood in order to reach this goal. Let us
mention two of them: first, one should make sure that there is no phase
transition at any finite value of $N$; for that reason it would be
 important to have some example where the exact explicit dependence on
 $N$ is known. Second, one would like to
understand the nature of the transition point which appears in the model
at a critical value $m_c$ of the mass parameter.
Indeed, looking at the induced
gauge theory, due to the fact that matter fields are in the adjoint
representation of the U(N) group (see below), one finds a super-
confining behaviour~\cite{ksw}, and the hope is to reach an ordinary
confining phase through the above mentioned
phase transition. Hence it would be interesting to study models which
are simple enough to be exactly solvable, but still having all the
 desired non trivial properties.

In this
letter we will give the exact solution for any value of N in
 the case of a d=1 compactified lattice made
of $S$ matter fields, for any value of $S$. This model fulfills the
 above requirements: it has a non trivial phase transition, and the
 solution can be  obtained by
using simple combinatoric properties of the permutation group, the key
trick being the reduction of the permutation group to its cyclic
representations. Moreover,
as one would expect from the definition of the model (see below),
our solution also describes
the vortex-free sector of the d=1 compactified bosonic string
{}~\cite{kle}.
In this context it is rather interesting to notice that the analytic
continuation  in the mass parameter from the strong to the weak coupling
phase provides the correct prescription to obtain the physical
properties of the upside-down oscillators from the standard matrix
oscillators ~\cite{bk}.

This letter is organized as follows: after a brief introduction on the
Kazakov-Migdal model (sect.2) , we give
in sect.3 the exact solution of the model.The solution is discussed in
sect. 4 which includes also  some concluding remark.

\vskip 1cm

{\bf 2. The Kazakov-Migdal model}
\vskip .3cm

The starting point of Kazakov-Migdal suggestion is to
induce the Yang-Mills interaction using massive scalar fields in the
adjoint representation of $U(N)$. The action they propose is
defined on a generic  d-dimensional lattice
and has the following form:

\eq
S = \sum_{x} N {\rm Tr} \bigl[
 m^{2} \phi^{2}(x) - \sum_{\mu} \phi(x)U(x,x+\mu)\phi(x+\mu)
U^{\dagger}(x,x+\mu)\bigr]
\en
where $\phi(x)$ is an Hermitian $N \times N$ matrix defined on the sites
$x$ of the lattice
and
 $U(x,x+\mu)$ is a Unitary $N \times N$ matrix, defined  on the links
$(x,x+\mu)$, and plays the role, as in the usual lattice
discretization of Yang-Mills theories, of the gauge field.
Integrating over the
scalar field $\phi$ one can induce an effective action for the gauge
field,
\eq
\int DU D\Phi exp (-S) \sim \int DU exp ( -S_{ind}[U])
\en
with:
\eq
S_{ind}[U] = - \frac{1}{2} \sum_{\Gamma} \frac{ |{\rm Tr} U[\Gamma]|^{2}}
{l[\Gamma] m^{2l[\Gamma]}}~~~,
\label{gauge}
\en
where $l[\Gamma]$ is the length of the loop $\Gamma$, $U[\Gamma]$ is
the ordered product of link matrices along $\Gamma$ and the summation
is over all closed loops.

In a similar way, integrating over the gauge fields
 one can induce an effective interaction for the scalar fields,
which turns out to be deeply related to the matrix approach to
 2D Quantum
Gravity and noncritical strings.

Integration can be achieved by using the well known  formula
first discovered by Harish-Chandra~\cite{hc}, rediscovered in the
context of matrix models by Itzykson and Zuber~\cite{iz} and fully
exploited by Mehta~\cite{me}:

\eq
I(\phi(x),\phi(y)) = \int D U \exp \left(
N \, tr \phi(x) U \phi(y) U^{\dagger}
\right) \propto \frac{\det_{ij} \exp(N \lambda_i(x)
\lambda_j(y) )}{\Delta(\lambda(x))
\Delta(\lambda(y))}
\label{izhc}
\en
where
$\lambda_i(x)$ are the eigenvalues of the matrix $\phi(x)$
\eq
\Delta(\lambda) = \prod_{i<j} (\lambda_i-\lambda_j)
\label{vandermond}
\en
is the Vandermonde determinant, and $(x,y)$ are nearest neighbour links
of the lattice.

The main difference between this approach and the usual one lies in the
fact that now, since the angular variables $U(x,x+\mu)$ are themselves
 degrees of freedom and their self-interaction term is explicitly
absent, we can integrate them  out safely, while the same is not
possible in
the usual description of 2d quantum gravity coupled with matter defined
on lattices with closed loops. It is exactly this kind of obstruction
which doesn't allow, in the context of the matrix approach,
a description of $d>1$ bosonic strings in terms of the eigenvalues
 only,  and which manifests itself in the case of the
d=1 compactified bosonic string as a  vortex contribution.
This means that in this last example the Kazakov-Migdal model should
exactly
correspond to the singlet, vortex free, solution of the compactified
bosonic string.

A further important feature of the Kazakov-Migdal model is the presence
 of a phase transition, which should occur at a finite, non-zero value
 $m_c$ of the mass parameter, between a strong coupling regime
 ($m> m_c$) and a weak coupling phase ($m< m_c$).
This transition was discussed in~\cite{km,migdal} in the context of the
induced matrix model (after integration on the gauge fields) and was
conjectured in~\cite{ksw} to be related (in the context of the induced
gauge theory, after integration on the matter fields) to the change from
ordinary to local confinement. We will show below that this same
transition in the d=1 case separates the upside-down oscillator phase
 from the standard matrix oscillator description of the d=1 bosonic
string.

\vskip 1cm

{\bf 3. Exact solution for a d=1 compactified lattice}
\vskip .3cm

\noindent
The partition function of the Kazakov-Migdal model defined on a 1d
lattice  with $S$ sites (labelled by $\alpha$) compactified on a circle
is:
\eq
Z~~ =
{}~\sum_{\alpha=1}^{S}
\int d \phi{(\alpha)}dU{(\alpha,\alpha+1)}~e^{-
N~~{\rm Tr}[m^2 \phi{(\alpha)^{2}}-\phi{(\alpha)}U{(\alpha,\alpha+1)}
\phi{(\alpha+1)}U{(\alpha,\alpha+1)^\dagger}]}~~.
\label{eq:c1}
\en

\noindent
The model (\ref{eq:c1}) is reduced , by integrating over the unitary
matrices on each link, to
\eq
Z~~= \int~\prod_{\alpha,i}~d\lambda_{i}^{(\alpha)}~e^{
-m^{2}N\sum_{\alpha,i}\lambda_{i}^{(\alpha)^{2}}}
\,\frac{1}{N^{SN(N-1)/2}}~\prod_{\alpha}\,\det
\left( e^{N\lambda_{i}^{(\alpha)
}\lambda_{j}^{(\alpha+1)}} \right)
\label{eq:c2}
\en
where $\lambda_i^{(\alpha)}$ is the $i^{th}$ eigenvalue of the matrix
$\phi(\alpha)$ .

It is easy to see that this expression can be rewritten as follows:
\eq
Z = \sum_{P_{\alpha}} (-1)^{P_{1}+ \cdots + P_{S}}
\int \prod_{\alpha,i} d\lambda_{i}^{(\alpha)} e^{-\frac{N}{2}
\sum_{i=1}^{N} \sum_{\alpha=1}^{S} [ a\lambda_{i}^{(\alpha)} - b(
P_{\alpha} \lambda^{(\alpha+1)})_{i} ]^{2}} \frac{1}{N^{SN(N-1)/2}}
\label{eq:c3}
\en
where the $P_{\alpha}$'s are $S$ independent permutations of $N$ objects
and we have defined  $2 m^{2} = a^{2} + b^{2} $  with  $ab = 1$ . Here
$a$ and $b$ are restricted to be real , hence $m >1$. The region with
$m<1$ ,where the integral (\ref{eq:c2}) is not defined, can be reached
by analytic continuation as described later.
Let us introduce the new variables of integration
\eqa
\hat\lambda_{i}^{(1)} & = & \lambda_{i}^{(1)} \nonumber \\
\hat\lambda_{i}^{(2)} & = & (P_{1} \lambda^{(2)})_{i} \nonumber \\
\hat\lambda_{i}^{(3)} & = & (P_{1}P_{2}\lambda^{(3)})_{i} \nonumber \\
            \vdots    &   &  \vdots  \nonumber \\
\hat\lambda_{i}^{(S)} & = & (P_{1}P_{2}\ldots P_{S-1} \lambda^{(S)})_{i}
\nonumber \\
\mbox{and define}~~~  P & = &  P_{1} \cdots P_{S} \nonumber
\ena
Then we have
\newpage
\eqa
Z & = & \frac{1}{N^{SN(N-1)/2}} (N!)^{S-1} \sum_{P} (-1)^{P}
\int \prod_{\alpha,i} d\hat\lambda_{i}^{(\alpha)}
{}~~\times \nonumber  \\
&  &
\exp  -\frac{N}{2} \sum_{i=1}^{N} \{ [a\hat\lambda_{i}^{(1)} -
b\hat\lambda_{i}^{(2)}]^{2} + [a\lhi^{(2)} - b \lhi^{(3)}]^{2}
+ \cdots + [a\lhi^{(S)} - b(P\hat\lambda^{(1)})_{i} ]^{2} \}
\nonumber \\
& & \mbox{}
\label{eq:c4}
\ena
It is natural now to change variables in the integral and define
\eq
\zeta_{i}^{(\alpha)}~~ = ~~ a \lhi^{(\alpha)} - b \lhi^{(\alpha+1)}
\en
where $ \lhi^{(S+1)} = (P\hat\lambda^{(1)})_{i} $.
Now we have to calculate the Jacobian.
This is a simple task if we take a cyclic permutation, say of order $r$,
namely the permutation: $1\to2\to3\to \cdots \to r \to1$ and if we order
the variables in the following way:
$$
\zeta^{(1)}_1,\zeta^{(2)}_1,\cdots\zeta^{(s)}_1,
\zeta^{(1)}_2,\zeta^{(2)}_2,\cdots\zeta^{(s)}_2,
\cdots\cdots
\zeta^{(1)}_r,\zeta^{(2)}_r,\cdots\zeta^{(s)}_r,
$$
then the Jacobian is given by:
\eq
\left| \frac{\partial \zeta}{\partial \hat\lambda}\right|
=\left|\matrix{a&-b& & & & & \cr
               &a&-b&&&&\cr
               &&a&-b&&&\cr
               &&&a&-b&&\cr
               &&&&.&.&\cr
               &&&&&a&-b\cr
               -b&&&&&&a}\right|
= a^{rS}-b^{rS}~~,
\label{eq:c6}
\en
where we have chosen $a>b$
due to the absolute value in the Jacobian.
Hence:
\eqa
a^2 & = & m^2+\sqrt{m^4-1} \nonumber \\
b^2 & = & m^2-\sqrt{m^4-1} \nonumber \\
\nonumber
\ena
\noindent
As each permutation $P$ can be decomposed into
products of cycles,  each integral at the r.h.s. of (\ref{eq:c4})
 decomposes
into the product of integrals corresponding to the cycles in
the decomposition of $P$.

Any permutation is characterized by a set of numbers $ \{ r_{1} ,
\cdots
,r_{j} , \cdots , r_{N} \} ~ ,~ r_{j} $ being the number of times that
the cycle of order j appears in the decomposition of P. Obviously one
has the condition
\eq
\sum_{j=1}^{N} j \, r_{j} ~ = ~ N
\label{eq:c7}
\en
Among the $N!$ permutations there are
 $ N! \, \prod_{j=1}^{N}(\frac{1}{j})^{r_{j}} \, \frac{1}{r_{j}!} $
which are characterized by the same set of numbers $\{ r_{i}\} $ .

\noindent
By putting everything together, we find for the partition function the
 following expression :
\eqa
Z & = & \frac{(N!)^{S}}{N^{SN(N-1)/2}}
\sum_{r_{1},\cdots , r_{N} }
\delta(\sum j \, r_{j} - N) (-1)^{\sum (j-1) r_{j}}
\prod_{j=1}^{N} \left( \frac{F_{j,S,N}}{j}\right)
^{r_{j}} \frac{1}{r_{j} !}
\nonumber \\
& = & \frac{(N!)^{S}}{N^{SN(N-1)/2}} \int_{0}^{2\pi} \frac{d\theta}{2
\pi}
\exp -iN \theta - \sum_{r=1}^{\infty} \frac{(-1)^{(r)}}{r} F_{r,S,N}
e^{i\theta r}
\label{eq:c8}
\ena
where $F_{r,S,N}$ is the integral corresponding to a cycle of order $r$:
\eq
F_{r,S,N} = \int \prod_{i=1}^{r} \prod_{\alpha=1}^{S}
d\lhi^{(\alpha)} e^{ - \frac{N}{2} \sum_{i=1}^{r}
\sum_{\alpha=1}^{S} ( a\lhi^{(\alpha)} -
b \lhi^{(\alpha + 1)} )^{2} }
\label{eq:c9}
\en
where as before $ \lhi^{(S+1)} = \hat\lambda_{i+1}
^{(1)} $ .

\noindent
But the r.h.s. of eq.(\ref{eq:c9}) is a gaussian integral in the
 $ \zeta_{i}^{(\alpha)} $ variables, hence:
\eq
F_{r,S,N} = \frac{1}{a^{rS} - b^{rS}}
\int \prod d\zeta_{i}^{(\alpha)} e^{-\frac{N}{2} \sum \zeta_{i}^{(\alpha
)^{2}} }
= \left( \frac{\pi}{N} \right)^{\frac{rS}{2}} \frac{1}
{a^{rS} - b^{rS}}
\label{eq:c10}
\en
By inserting (\ref{eq:c10}) into the expression (\ref{eq:c8}) for Z we
obtain
\eq
Z = \frac{(N!)^{S}}{N^{SN(N-1)/2}} \int_{0}^{2\pi} \frac{d \theta}{2\pi}
e^{-iN\theta} \prod_{K=0}^{\infty} \left(
1 + e^{i\theta} \left( \frac{\pi}{Na^{2}} \right)^{S/2} a^{-2SK} \right)
\label{eq:c11}
\en
In this
infinite product we recognize the grand-canonical partition function of
a
set of fermions where $A\equiv e^{i\theta} \left( \frac{\pi}{Na^{2}}
\right)^{S/2}$ plays the role of the fugacity. Then
the only effect of the integral over $\theta$ is to select in the
product $ \prod_{K=0}^{\infty} ( 1 + AX^
{SK} ) $
the coefficient of the $ A^{N} $ term. Such
coefficient is
\eq
\sum_{K_{1} < K_{2} < \cdots < K_{N}} X^{SK_{1}} \, X^{SK_{2}} \ldots
X^{SK_{N}} = \frac{1}{X^{NS}} \prod_{K=1}^{N}
\frac{X^{SK}}{1 - X^{SK}}
\label{eq:c12}
\en
So we find the final expression for the partition function
\eq
Z~ =~\frac{1}{2\pi} \frac{(N!)^{S}}{N^{SN(N-1)/2}} \left(
\frac{\pi}{N} \right)^{\frac{NS}{2}}
\, \prod_{K=1}^{N} \frac{(a^{-2S})^{N^{2}/2}}{1 - (a^{-2S})^{K}}
\label{eq:c13}
\en
\newpage
{\bf 4.  Discussion of the results and concluding remarks}
\vskip 0.3cm
Apart from some irrelevant factors the partition function (\ref{eq:c13})
 is of the form
\eq
 Z^{(N)}(q)~=~\frac{q^{N^2/2}}{(1-q)(1-q^2)\cdots (1-q^N)}
\label{d1}
\en
with $ q= a^{-2S} $ .

The same partition function  has been obtained in a completely different
fashion by Boulatov and Kazakov in ref ~\cite{bk} as the one describing
 the singlet (vortex free) part of the partition function for a 1d
 string.
This is not surprising as the singlet is obtained by integrating over
the residual angular variables which play therefore the role of gauge
variables.
The crucial difference is that the theory considered in ~\cite{bk} has a
 continuous
compactified target space. As a consequence the argument $ q = a^{-2S} $
in (\ref{d1})is identified in ~\cite{bk} with $ q = e^{-\beta\omega} $
 where  $\beta $ is the length of the string and $ \omega$
 is the frequency of the
oscillators. A rescaling of time changes both $ \beta$ and $ \omega$
 in such a way to keep this product constant.
The most convenient way to formulate such scaling in our case is to
define
\eq
a^{2} = e^{\varphi}
\label{d2}
\en
so that
\eq
m^{2} = \cosh \varphi
\label{d3}
\en
 The partition function (\ref{eq:c13}) is then left invariant by the
following rescaling :
\eq
S \to  S' ~~~~~~~~~~~~~ \varphi \to \frac{S}{S'} \varphi
\label{d4}
\en
which for the mass of the scalar implies
\eq
m^{2} \equiv \cosh \varphi \to m'^{2} \equiv
\cosh( \frac{S}{S'} \varphi )
\label{d5}
\en
With an infinite rescaling $ (S' \to \infty ) $
one should recover the continuum theory of ref. ~\cite{bk}.
In fact it can be easily checked that the action in (\ref{eq:c1})
becomes in such limit  :
\eq
S_{cont} = N {\rm Tr} \int_{0}^{\beta} dt \left[ \frac{1}{2} (D
\hat{\phi} )^{2} + \frac{1}{2} \omega^{2} \hat{\phi}^{2} \right]
\label{d6}
\en
where $ t = \frac{\alpha}{S'} \beta$ , $ \hat{\phi}
= \sqrt{ \frac{\beta}{S'} } \phi $ and
\eq
\beta^{2} \omega^{2} = 2 S'^{2} (m'^{2} - 1 ) =
2 S'^{2} [ \cosh(\frac{S}{S'} \varphi ) - 1]
= \varphi^{2} + O(\frac{1}{S'^{2}} )
\label{d7}
\en
It is apparent from (\ref{d7}) that the continuum limit always
correspond to the critical point $ m^{2} = 1 $, although it leads to two
completely different phases according to whether the point
 $ m^{2} =1 $ is
reached from above or from below, the former corresponding to a real and
the latter to an imaginary frequency $ \omega $ .

\noindent
The fact that , except for a rescaling of $a$, the same partition
 function is obtained irrespective of the value of $S$ is quite
 remarkable and tells us that the whole information about the partition
 function is
already contained in the simplest case $ S = 1 $
where the lattice is reduced to just one site and one link.
Such drastic reducibility in the number of degrees of freedom seems
to denote that all relevant quantities, such as
for instance the vacuum density of the eigenvalues of the scalar field ,
are space independent.

The calculation leading to eq. (\ref{eq:c13}) has been done in the
regime $ m > 1 $. In such regime the quadratic potential is stable
, $ a $ and $ b$ are real and all integrals are well defined. In the
weak coupling regime ($ m < 1 $) the quadratic potential is unstable
and such instability manifests itself in the divergence of the integrals
over the eigenvalues. It is easy to see for instance that for $ a $ and
$ b $ on the unit circle the integral in (\ref{eq:c9}) is divergent.

\noindent
The natural way out is to define the partition function for $ m < 1 $
as the analytic continuation from $ m > 1 $.
In terms of the variable $ q = a^{-2S} $ it means an analytic
continuation from the real axis with $ q < 1 $ to the unit circle.

\noindent
As discussed above, in  the continuum theory we have $ m \to m_{c} = 1 $
 and two different phases originate corresponding to real (resp.
 imaginary) frequency oscillators if $ m_{c}$ is approached from above
 (resp. below).
It was argued in ref. ~\cite{bk} that the physical properties of the
oscillators with imaginary frequencies - the so called upside-down
oscillators - can be obtained from the ones of ordinary oscillators by
the replacement $ \omega \to i\omega $ .
It was shown this to be consistent with the introduction of an SU(N)
invariant cutoff at large $ \lambda$ 's.

\noindent
In the discrete theory on the other hand it is possible to analytically
continue from the strong to the weak coupling regime and then
 perform the continuum limit. In this way one recovers the correct
prescription for the upside-down oscillators, namely that their
properties  are obtained via the substitution $ \omega \to i\omega $ .

Finally we want to remark that it is very simple in this theory to
integrate over the matrix fields $\phi $ and obtain the partition
function as a function of the gauge fields only. By taking advantage of
the fact that the partition function is independent from S we can do the
calculation for $ S = 1 $.
Gauge invariance can be used to choose the unitary matrix $U$ to be
diagonal and the integral over each matrix element of the matrix field
$\phi$ is gaussian. The result is then :
\eqa
Z(\beta) & = &
\int_{0}^{2\pi} \prod_{k=1}{N} \frac{d\theta_{k}}{2\pi}
\vert\Delta(e^{i\theta})\vert^2 \left[ \frac{1}{m^{2}-1} \right]
^{N/2} \prod_{i<j} \frac{1}{2[m^{2} - \cos(\theta_{i}-\theta_{j})]}
\nonumber
\\
 & = & \int_0^{2\pi}\prod_{k=1}^N\frac{d\theta_k}{2\pi}
\vert \Delta(e^{i\theta})\vert^2
\prod_{k,m=1}^{N}\left[
\frac{q^{1/2}}{1-qe^{i(\theta_{k}-\theta_{m})}}
\right]
\label{eq:c14}
\ena
where, for $S=1$, $q=1/a^2$ and the $\theta_i$'s are the invariant
angles of the $U(N)$ matrix.
In order to obtain the expression for arbitrary S it is sufficient to
 replace $q$ with $a^{-2S}$ and interpret the $\theta_{i}$'s as the
 invariant angles of the product of the unitary matrices over the
 plaquette.
It should be noticed that eq. (\ref{eq:c14}) is much more suitable than
eq. (3) to study the weak and strong coupling behaviour of the induced
gauge theory since it can be analytically continued from the strong to
the weak coupling regime.

\noindent
Eq. (\ref{eq:c14}) was also derived, in the continuum limit, in ref.
 ~\cite{bk} (see eq. (4.34))  by  performing the gaussian
integral over $ \hat{\phi}(t)$ and it was used as an intermediate step
 in deriving eq. (\ref{d1}).
\vskip 1cm
\noindent {\bf Note added}
\par
After completing this paper we have become aware of a paper by S. Dalley
entitled "The Weingarten model \`a la Polyakov" and published on Mod.
Phys. Lett. \bf A7 \rm (1992) 1651.  Induced gauge theories on a lattice
 are considered  there  in the context of a complex
matrix model ; in particular the case $D=1$ , that has the same physical
content as the corresponding Kazakov-Migdal model, is studied in detail.

\vskip 1cm
\noindent {\bf Acknowledgments}
\par
We thank F. Gliozzi  for many enlightening discussions


\vfill
\newpage
\begin{thebibliography}{99}

\bibitem{km} V.A. Kazakov and A.A. Migdal, "Induced QCD at Large
 $N$", preprint PUPT - 1322, LPTENS -92/15, May 1992
\bibitem{migdal} A.A. Migdal, "Exact Solution of Induced Lattice
 Gauge Theory at large
 $N$", preprint PUPT - 1323,Revised, June 1992
\bibitem{ksw}
I.I. Kogan, G.W. Semenoff and N. Weiss,
 "Induced QCD and Hidden Local Z$_{\bf N}$ Symmetry", preprint
 UBCTP 92-022  June 1992
\bibitem{kle} D.J.Gross and I.R.Klebanov, Nucl. Phys. \bf B344 \rm
(1990) 475 and \bf B354 \rm (1991) 459.
\bibitem{bk} D. Boulatov and V. Kazakov, "One-dimensional string theory
with vortices as the upside-down matrix oscillator", preprint LPTENS 91/
24, KUNS 1094 HE(TH) 91/14, August 1991
\bibitem{iz} C.Itzykson and J.B.Zuber, J.Math.Phys. 21 (1980) 411.
\bibitem{hc} Harish-Chandra, Amer.J.Math. 79 (1957) 87.
\bibitem{me} M.L.Mehta, Comm.Math.Phys. 79 (1981) 327.

\end{thebibliography}
\end{document}